\let\pdfoutput\defined
\documentclass{PoS}

\title{%
\vspace*{-1cm}
\begin{minipage}{\textwidth}
\begin{flushright}
\texttt{\footnotesize
PoS(Lattice 2011)142\\%
}
\end{flushright}
\end{minipage}\\[15pt]
SU(2) chiral perturbation theory low-energy constants from staggered 2+1 flavor simulations}

\ShortTitle{SU(2) chPT low-energy constants from staggered 2+1 flavor simulations}

\author{\speaker{Enno~E.~Scholz}$^a$, Szabolcs~Bors\'anyi$^b$, Stephan~D\"urr$^{b,c}$, Zolt\'an~Fodor$^{b,c,d}$, Sandor~D.~Katz$^{b,d}$, Stefan~Krieg$^{b,c}$, Andreas~Sch\"afer$^a$, and Kalman~K.~Szabo$^b$\\
\llap{$^a$}Universit\"at Regensburg, Universit\"atsstr.~31, D-93053 Regensburg, Germany\\
\llap{$^b$}Bergische Universit\"at Wuppertal, Gau\ss{}str.~20, D-42119 Wuppertal, Germany\\
\llap{$^c$}J\"ulich Supercomputing Centre, Forschungszentrum J\"ulich, D-52425 J\"ulich, Germany\\
\llap{$^d$}Institute for Theoretical Physics, E\"otv\"os University, H-1117 Budapest, Hungary\\
E-mail: \email{enno.scholz@physik.uni-regensburg.de}%
, \email{durr@itp.unibe.ch}
}

\abstract{%
We measure the pion mass and decay constant on ensembles generated by the Wuppertal-Budapest Collaboration, and extract the NLO low-energy constants $\bar{l}_3$ and $\bar{l}_4$ of SU(2) chiral perturbation theory. The data are generated in 2+1 flavor simulations with Symanzik glue and 2-fold stout-smeared staggered fermions, with pion masses varying from 135 MeV to 400 MeV, lattice scales between 0.7 GeV and 2.0 GeV, and $m_{\rm s}$ kept at its physical value. Furthermore, by excluding the lightest mass points, we are able to test the reliability of SU(2) chPT as a tool to extrapolate towards the physical point from higher pion masses.%
}

\FullConference{XXIX International Symposium on Lattice Field Theory \\
		 July 10--16, 2011\\
		 Squaw Valley, Lake Tahoe, California}

\newlength{\closercaption}
\newlength{\afterTable}
\newlength{\afterFigure}
\newlength{\closersection} 
\begin{document}
%%%%%%%%%%%%%%%%%%%%%%%%%%%%%%%%%%%%%%%%%%%%%%%%%%%%%%%%%%%%%%%%%%%%%%%%%%%%%%%

%%%%%%%%%%%%%%%%%%%%%%%%%%%%%%%%%%%%%%%%%%%%%%%%%%%%%%%%%%%%%%%%%%%%%%%%%%%%%%%
\section{Introduction}
\label{sec:intro}
\vspace*{\closersection}

Chiral perturbation theory (chPT) \cite{Gasser:1983yg,Gasser:1984gg} is a widely used tool in many phenomenological applications and also helpful to guide an extrapolation to lighter quark masses in lattice-QCD simulations. Here we will report on a determination of the NLO low-energy constants (LECs) $\bar{l}_3$ and $\bar{l}_4$ which appear in the light quark mass dependence of the pseudo-scalar meson masses and decay constants in SU(2) chPT.

We analyze configurations generated by the Wuppertal-Budapest Collaboration \cite{Aoki:2006we,Aoki:2006br,Aoki:2005vt,Aoki:2009sc,Borsanyi:2010bp,Borsanyi:2010cj} using the Symanzik glue and 2-fold stout-smeared staggered fermion action for a 2+1 flavor QCD-simulation. The mass of the single flavor has been kept at the value of the physical strange quark mass, whereas the two degenerate lighter quark masses have been varied such that light meson masses in the range of 135 to 440 MeV were simulated. The simulations were performed at five different gauge couplings $\beta$, resulting in lattice scales between 0.7 and 2.0 GeV (see next section for details on how the scale has been determined). Table~\ref{tab:ensembles} summarizes some of the parameters of the simulations.

The 2-fold stout-smeared version of the staggered quark action has been proven to be advantageous \cite{Borsanyi:2010bp} in reducing the inevitable taste-breaking of staggered fermion formulations. Therefore, in this work we only consider the pseudo-scalar mesons with taste matrix $\gamma_5$ when measuring meson masses or decay constants. Details of the computation of these quantities will be reported in a forthcoming publication.

%%%%%%%%%%%%%%%%%%%%%%%%%%%%%%%%%%%%%
% table: ensembles
%%%%%%%%%%%%%%%%%%%%%%%%%%%%%%%%%%%%%
\begin{table}[b]
\begin{center}
\begin{tabular}{cccc}
$\beta$ & $1/a$ [GeV] &  $m_l/m_l^{\rm phys}$ (approx.) & $(L/a)^3\times(T/a)$\\\hline
3.45 & 0.69 & 1.0, 3.0, 5.0, 7.0, 9.0 & $24^3\times32$ -- $12^3\times28$\\
3.55 & 0.91 & 1.0, 3.5, 5.0, 7.0, 9.0 & $24^3\times32$ -- $12^3\times28$\\
3.67 & 1.31 & 1.0, 4.0, 6.0, 7.5, 9.5 & $32^3\times48$ -- $14^3\times32$\\
3.75 & 1.62 & 1.0, 4.0, 6.0, 8.0, 10.0 & $40^3\times64$ -- $16^3\times32$\\
3.85 & 2.04 & 1.0, 1.4, 2.0, 4.0, 6.0, 8.0, 10.0 & $48^3\times64$ -- $24^3\times
48$
\end{tabular}
\end{center}
\vspace*{\closercaption}
\caption{Simulated lattice ensembles: gauge coupling $\beta$, lattice spacing $1/a$, simulated quark masses $m_l$, and range of lattice sizes.}
\label{tab:ensembles}
\vspace*{\afterTable}
\end{table}
%%%%%%%%%%%%%%%%%%%%%%%%%%%%%%%%%%%%%

%%%%%%%%%%%%%%%%%%%%%%%%%%%%%%%%%%%%%%%%%%%%%%%%%%%%%%%%%%%%%%%%%%%%%%%%%%%%%%%
\section{Scale setting and physical quark masses}
\label{sec:scale_mud}
\vspace*{\closersection}

\begin{figure}[t!]
\begin{center}
\includegraphics*[angle=-90,width=.49\textwidth]{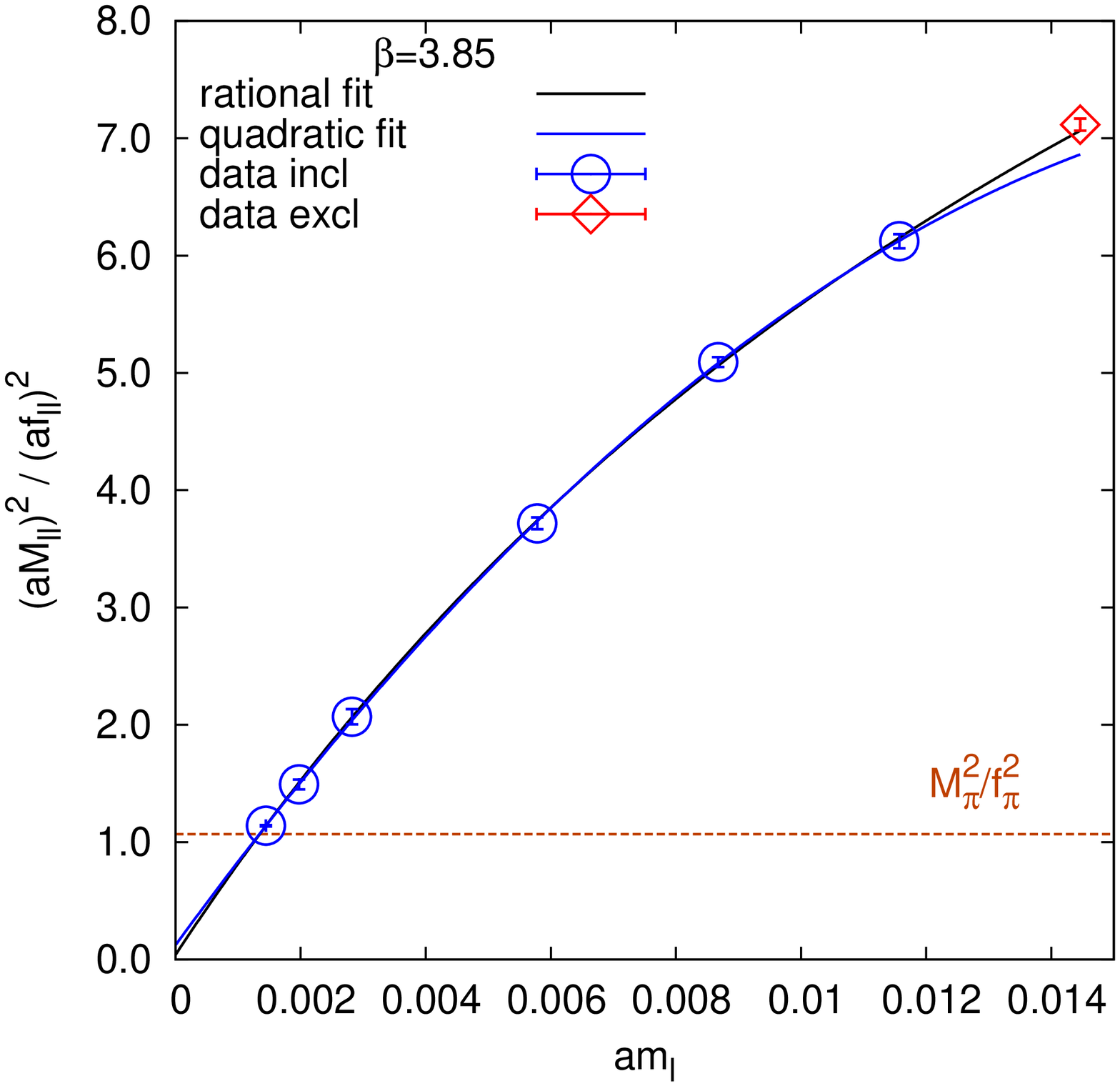}%
\includegraphics*[angle=-90,width=.49\textwidth]{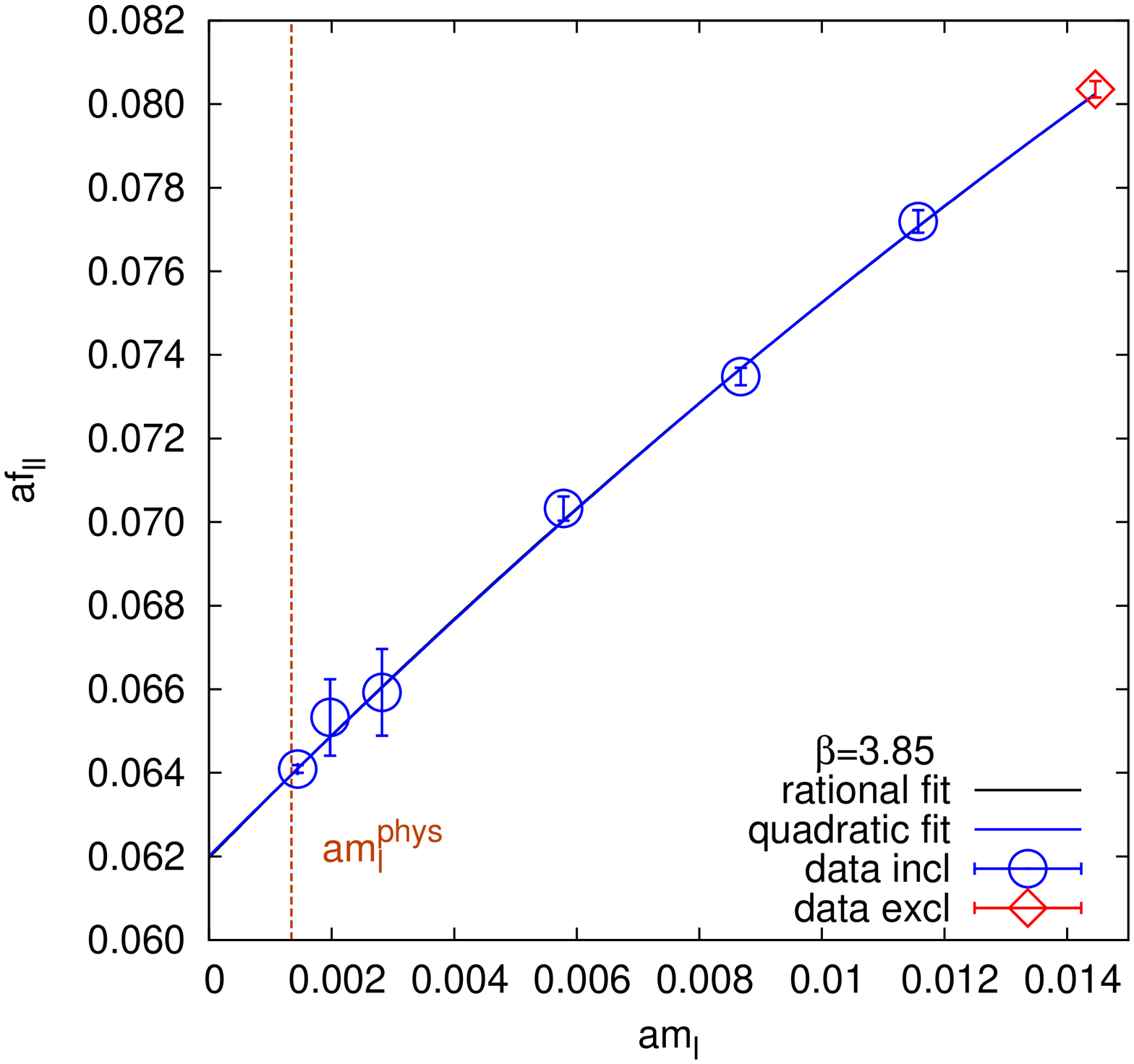}%
\end{center}
\vspace*{\closercaption}
\caption{{\it Left panel:} ratio $(aM_{ll})^2/(af_{ll})^2$ extrapolated to $M_\pi^2/f_\pi^2=1.06846$ to obtain $am_l^{\rm phys}$, {\it right panel:} $af_{ll}$ extrapolated to $am_l^{\rm phys}$ to obtain $1/a$; both at $\beta=3.85$.}
\label{fig:set_mud_scale}
\end{figure}

To set the scale at each simulated gauge coupling $\beta$ and identify the physical point, i.e.\ the average up/down quark mass $m_l^{\rm phys}=(m_{\rm u}+m_{\rm d})/2$ corresponding to a pion in the isospin limit with an estimated mass of $M_\pi=134.8\,{\rm MeV}$ \cite{Colangelo:2010et}, we use a two-step procedure. First, we extrapolate the ratio $(aM_{ll})^2/(af_{ll})^2$ of the squared meson masses and decay constants to $M_\pi^2/f_\pi^2 = (134.8\,{\rm MeV} / 130.41\,{\rm MeV})^2 = 1.06846$, where we also used the PDG-value $f_\pi=130.41\,{\rm MeV}$ \cite{Nakamura:2010zzi}. In that way $am_l^{\rm phys}$ is obtained. In the second step, we extrapolate $af_{ll}$ to this quark mass value and obtain the lattice scale with the help of the PDG-value for $f_\pi$. For the extrapolation we used two different ans\"atze: a quadratic and a rational (linear in numerator and denominator) fit form. An example of these extrapolations is shown for the ensembles at $\beta=3.85$ in Fig.~\ref{fig:set_mud_scale}. There, like for all other $\beta$-values as well, the heaviest quark mass point has been excluded, resulting in a fit range of approx.\ $am_{l}/am_l^{\rm phys} \leq 8.0$ (corresponding to $M_{ll}\leq390\,{\rm MeV}$). We stress that here, like in the chiral fits to be discussed below, the data has been corrected for finite volume effects beforehand, by means of using the two- and three-loop resummed formulae of \cite{Colangelo:2005gd} for the pion decay constants and masses, respectively. Our spatial lattice volumes $L^3$ are in the range $(4.7\,{\rm fm})^3$ -- $(6.8\,{\rm fm})^3$ with a minimal $M_{ll}L\approx 3.2$, ensuring that the finite volume corrections within our fit ranges are at most at the order of 1 per cent for the decay constants and even less for the meson masses.

By fixing $1/a$ and $am_l^{\rm phys}$ in the way described above, the meson masses and decay constants show no discretization effects at all directly at the physical point and we can assume those effects to be small (since of higher order in the quark masses and/or lattice spacing) in the vicinity of the physical point, i.e.\ in the mass range covered by our fits. Such discretization effects, of course, are present in other observables, which are not considered in this work.

%%%%%%%%%%%%%%%%%%%%%%%%%%%%%%%%%%%%%%%%%%%%%%%%%%%%%%%
% ratio and fPi to set mud and scale
%%%%%%%%%%%%%%%%%%%%%%%%%%%%%%%%%%%%%%%%%%%%%%%%%%%%%%%
\vspace*{\afterFigure}
%%%%%%%%%%%%%%%%%%%%%%%%%%%%%%%%%%%%%%%%%%%%%%%%%%%%%%%

%%%%%%%%%%%%%%%%%%%%%%%%%%%%%%%%%%%%%%%%%%%%%%%%%%%%%%%%%%%%%%%%%%%%%%%%%%%%%%%
\section{Fits to NLO SU(2) chPT}
\label{subsec:fits.nlo}
\vspace*{\closersection}

The quark mass dependence of the finite-volume corrected data for the meson masses and decay constants is fitted simultaneously at different $\beta$-values using the NLO-SU(2) chPT formulae
%%%%
\begin{eqnarray}
M_{ll}^2 &=& \left(\frac{1}{a}\right)^2 (aM_{ll})^2 \;=\; \chi_l\,\left[1\,+\,\frac{\chi_l}{16\pi^2 f^2}\log\frac{\chi_l}{\Lambda_3^2}\right]\,, \\
f_{ll} &=& \left(\frac1{a}\right) (af_{ll}) \;=\; f\left[1\,-\,\frac{\chi_l}{8\pi^2f^2}\log\frac{\chi_l}{\Lambda_4^2}\right]\,,\\
\chi_l &=& 2B\,m_l \;=\; (2Bm_l^{\rm phys})\,\frac{am_l}{am_l^{\rm phys}}\,,
\end{eqnarray}
%%%%
where we made use of the already determined $1/a$ and $am_l^{\rm phys}$ to scale the quark masses and the meson masses and decay constants measured in lattice units. This fit has four free parameters: two NLO low-energy scales $\Lambda_3$, $\Lambda_4$, the decay constant in the SU(2) chiral limit $f$ and the renormalization scheme-independent combination $(2Bm_l^{\rm phys})$ of the LO low-energy constant $B$ and the physical quark mass $m_l^{\rm phys}$.

We would like to point out that the chiral fit formulae do not include any taste breaking effects, i.e., we did not use staggered chPT. This seems justified to us, since we are only considering $\gamma_5$-taste mesons as mentioned above and use these to define our scaling trajectory at the physical point. In other words, since the meson mass and decay constant at the physical point were used to set the quark masses and lattice scales, no discretization or taste breaking effects are present in the chPT formulae for $M_{ll}^2$ and $f_{ll}$ as discussed above. Furthermore, taste breaking effects are reduced anyway by the choice of the fermion action as mentioned above.

\begin{figure}[ht!]
\begin{center}
%%%%
\includegraphics*[angle=-90,width=.49\textwidth]{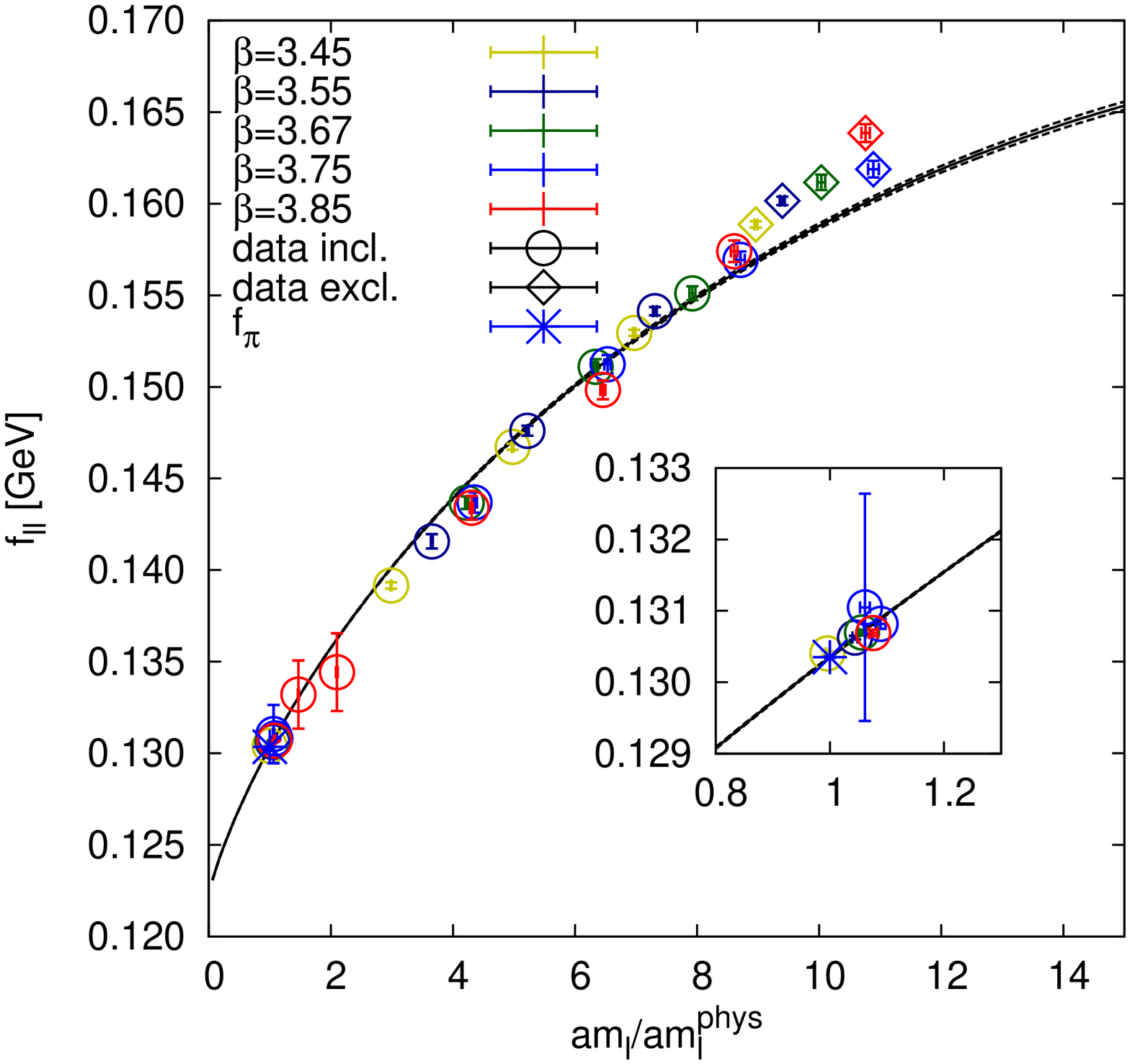}%
\includegraphics*[angle=-90,width=.49\textwidth]{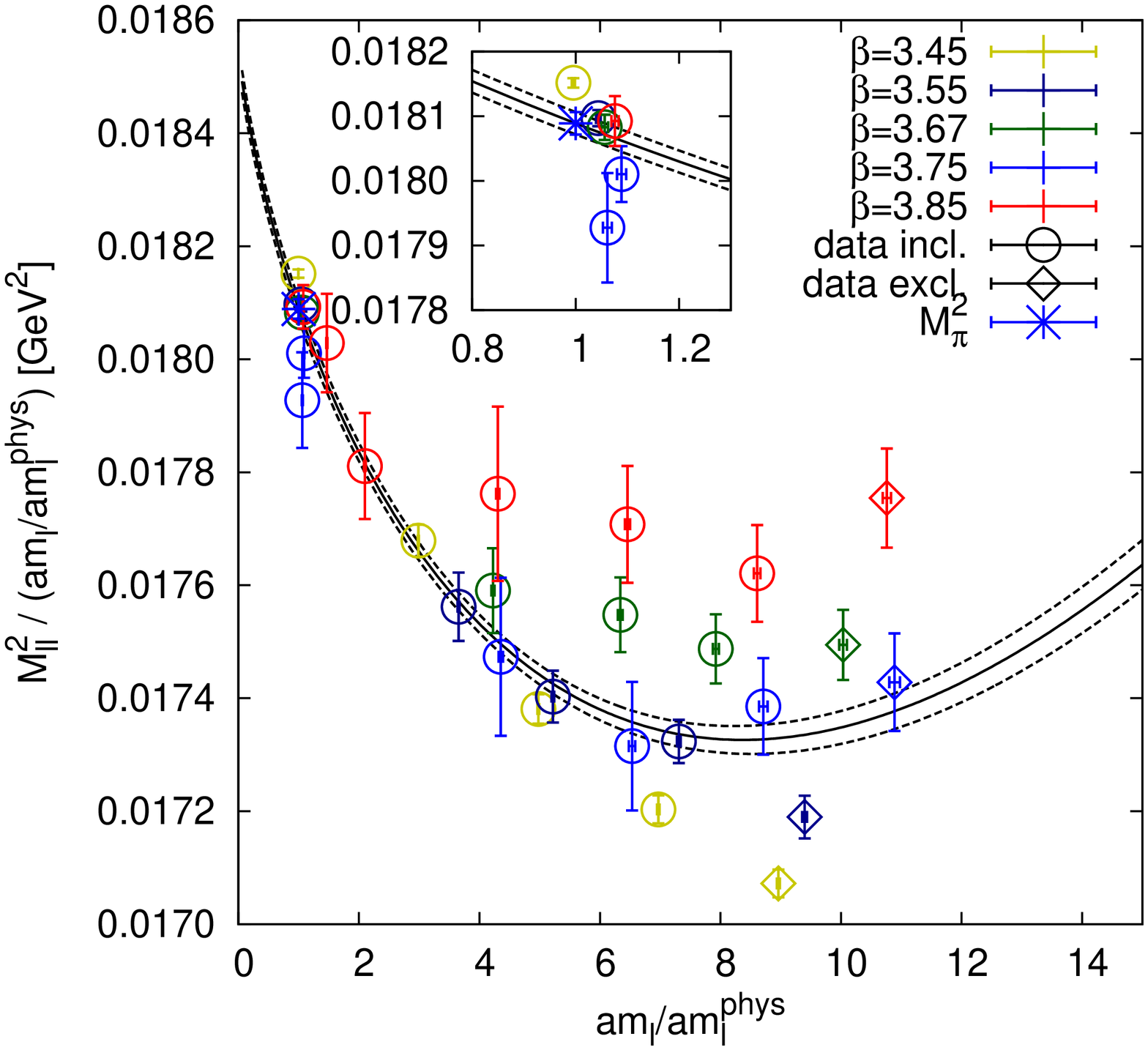}\\
%%%%
\includegraphics*[angle=-90,width=.49\textwidth]{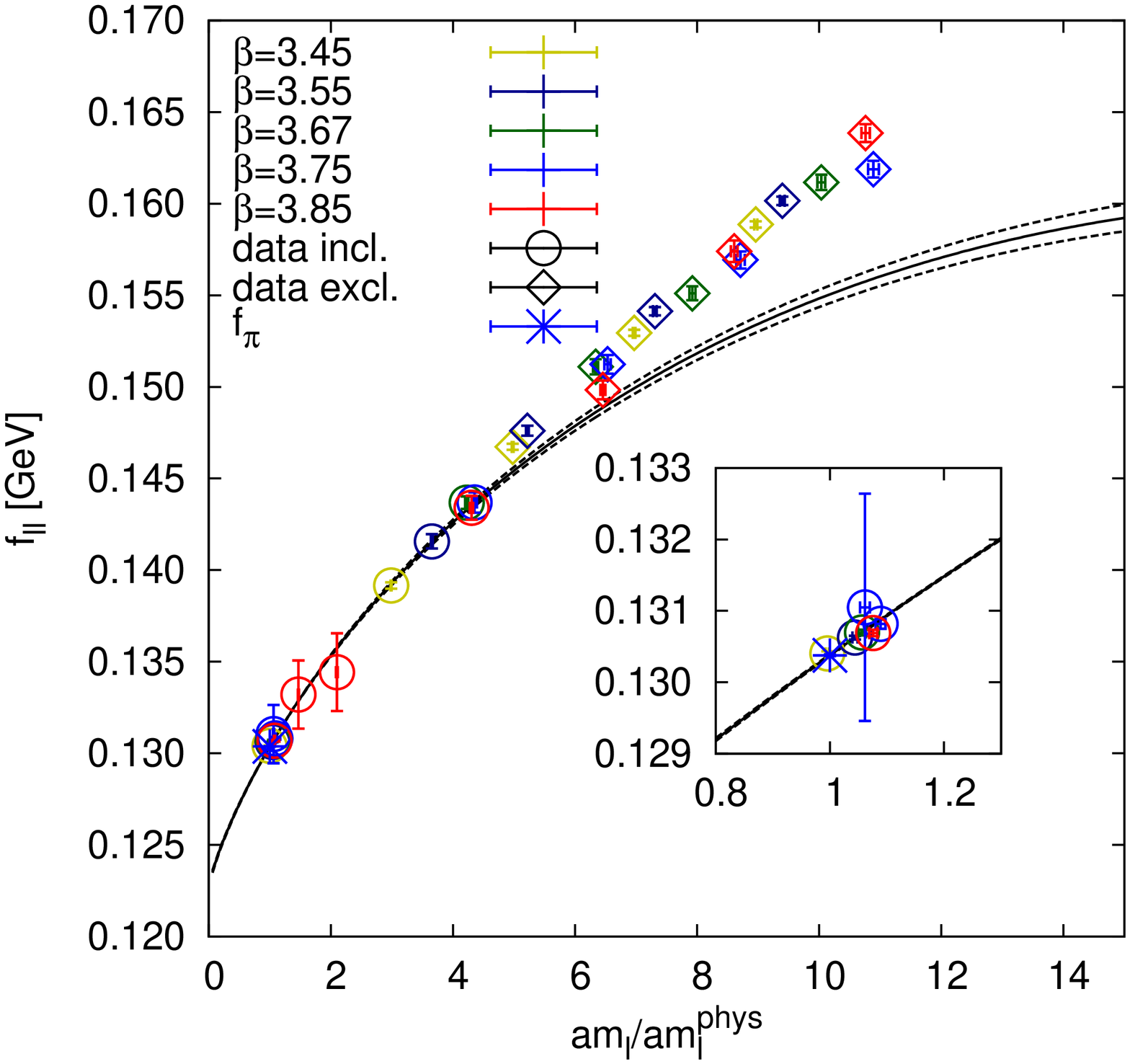}%
\includegraphics*[angle=-90,width=.49\textwidth]{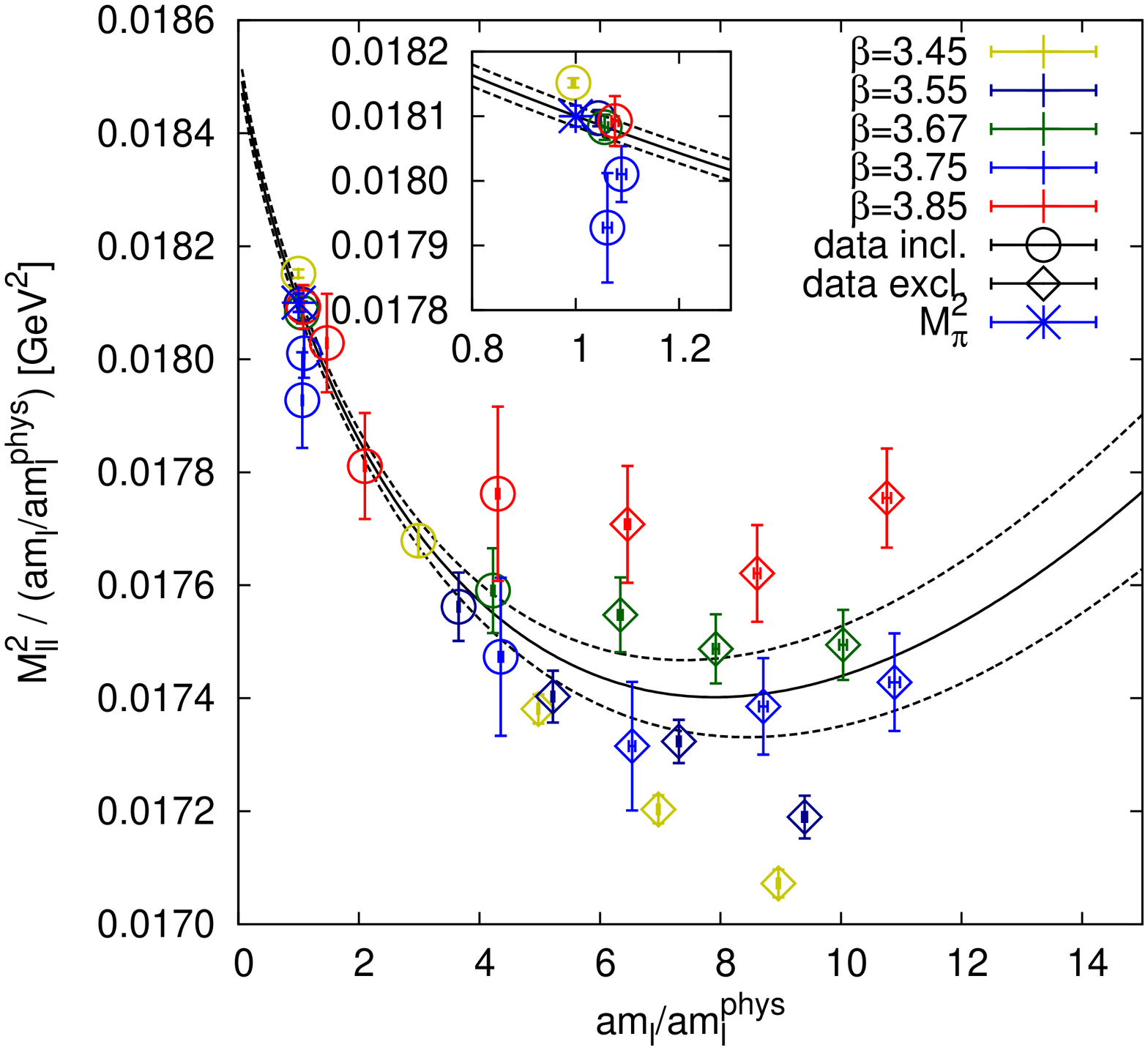}\\
%%%%
\includegraphics*[angle=-90,width=.49\textwidth]{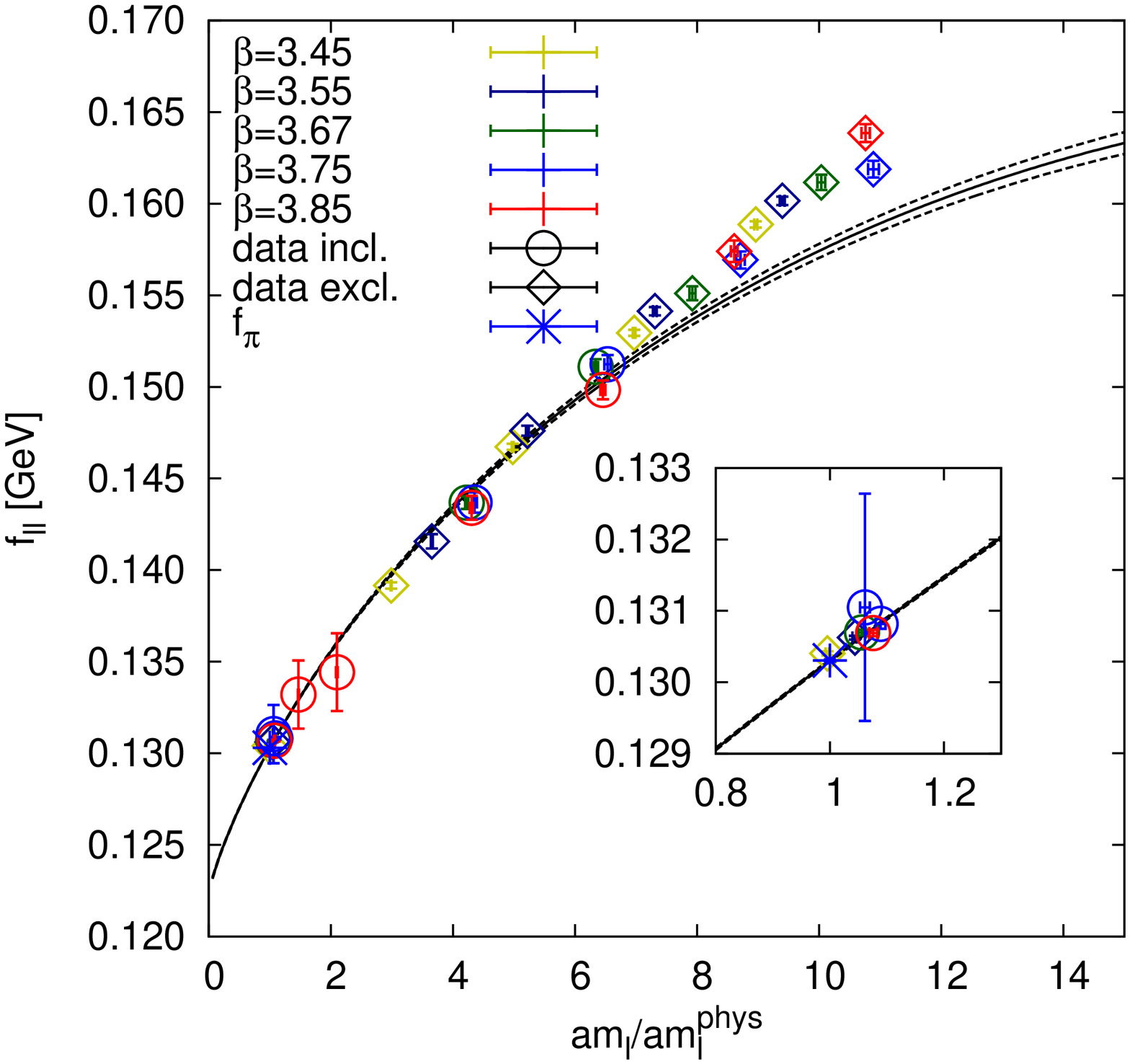}%
\includegraphics*[angle=-90,width=.49\textwidth]{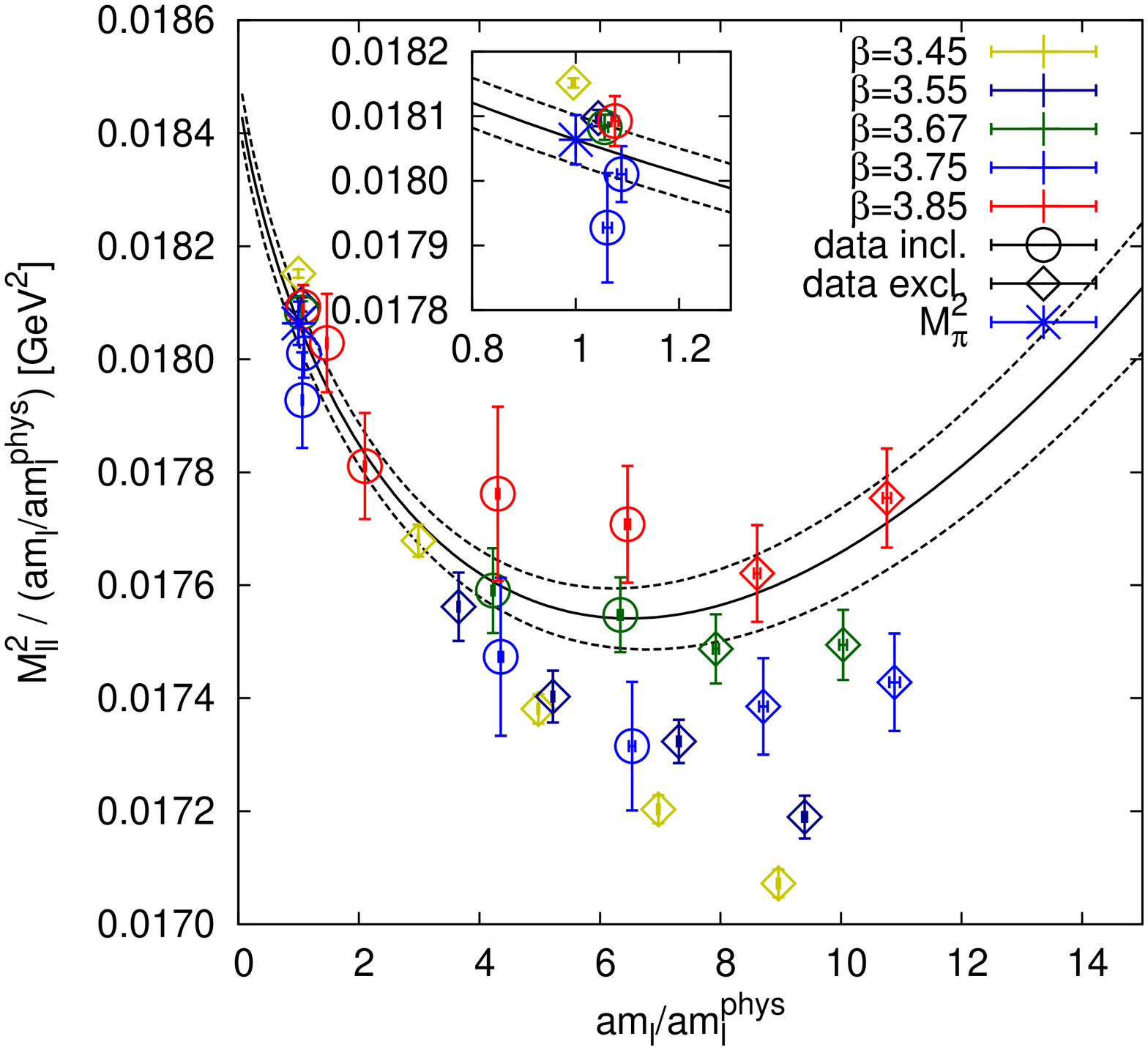}\\
%%%%
\end{center}
\vspace*{\closercaption}
\caption{Combined NLO SU(2) chPT fits with various fit ranges. The {\it left panels} show the decay constants $f_{ll}$, the {\it right panels} the squared meson masses $M_{ll}^2$ divided by the quark mass ratio $am_l/am_l^{\rm phys}$. The fit ranges are: {\it top:} all $\beta$, $135\,{\rm  MeV} \leq M_{ll} \leq 390\,{\rm MeV}$, {\it middle:} all $\beta$, $135\,{\rm  MeV} \leq M_{ll} \leq 275\,{\rm MeV}$, {\it bottom:} only $1/a\geq 1.35\,{\rm GeV}$, $135\,{\rm  MeV} \leq M_{ll} \leq 340\,{\rm MeV}$.}
\label{fig:fits_NLO}
\end{figure}

The top panels of Fig.~\ref{fig:fits_NLO} show the combined fits including the data at all lattice spacings and for meson masses in the range of 135 to 390 MeV. (Here and for all following plots we mark data points included in the fit by circles, while those not included in the fit are marked by diamonds.) As one can already see by eye, the description of the data by the fit is not satisfactory, resulting in a $\chi^2/{\rm d.o.f.}\approx 4.3$. As expected, the fit quality measured, e.g., by $\chi^2/{\rm d.o.f.}$ improves continuously when reducing the upper bound of the meson mass range. The middle panel of Fig.~\ref{fig:fits_NLO} shows the fit to all meson masses in the range $135\,{\rm MeV}\leq M_{ll}\leq275\,{\rm MeV}$ giving an acceptable $\chi^2/{\rm d.o.f.}\approx 1.0$. A similar improvement can be achieved by excluding the two coarsest lattice ensembles from the fit, i.e., limiting $1/a\geq 1.3\,{\rm GeV}$. The bottom panel of Fig.~\ref{fig:fits_NLO} shows an example of such a fit with  $135\,{\rm MeV}\leq M_{ll}\leq 340\,{\rm MeV}$ resulting in  $\chi^2/{\rm d.o.f.}\approx 1.7$ (this number has to be compared to $\chi^2/{\rm d.o.f.}\approx 2.6$ for the same mass range and using all $\beta$). Applying both kinds of cuts, i.e.\ $135\,{\rm MeV}\leq M_{ll}\leq275\,{\rm MeV}$ and $1/a\geq1.3\,{\rm GeV}$, eventually gives a $\chi^2/{\rm d.o.f.} \approx 0.8$.

%%%%%%%%%%%%%%%%%%%%%%%%%%%%%%%%%%%%%%%%%%%%%%%%%%%%%%%
% various NLO-fits (for each fPi and mPiSqr/ml): 
%
% * all beta, 135-390 MeV
%
% * all beta, 135-275 MeV
%
% * 1/a >= 1.35 GeV, 135-340 MeV
%
%%%%%%%%%%%%%%%%%%%%%%%%%%%%%%%%%%%%%%%%%%%%%%%%%%%%%%%
\vspace*{\afterFigure}
%%%%%%%%%%%%%%%%%%%%%%%%%%%%%%%%%%%%%%%%%%%%%%%%%%%%%%%

\begin{figure}[t!]
\begin{center}
\includegraphics*[width=.33\textwidth]{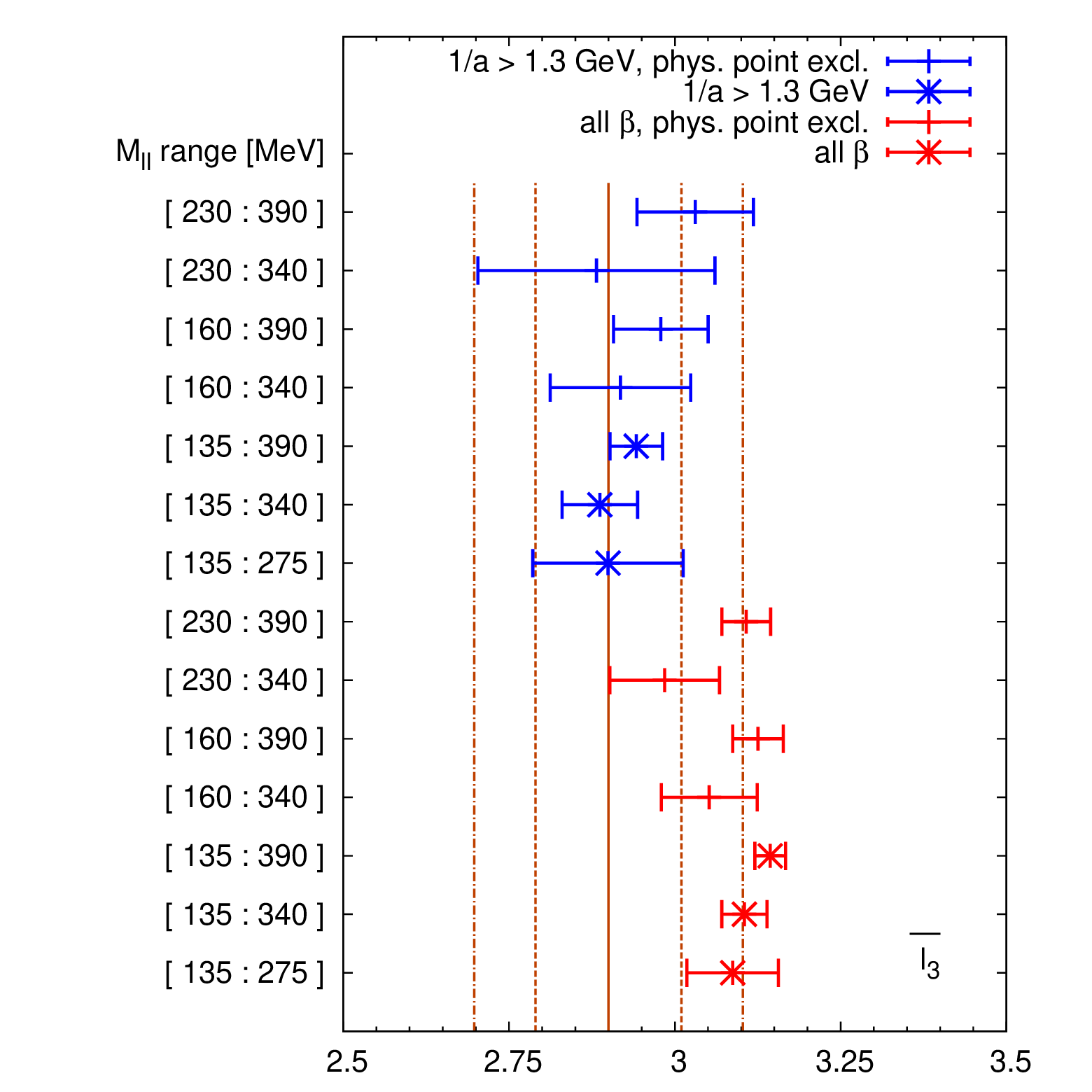}%
\includegraphics*[width=.33\textwidth]{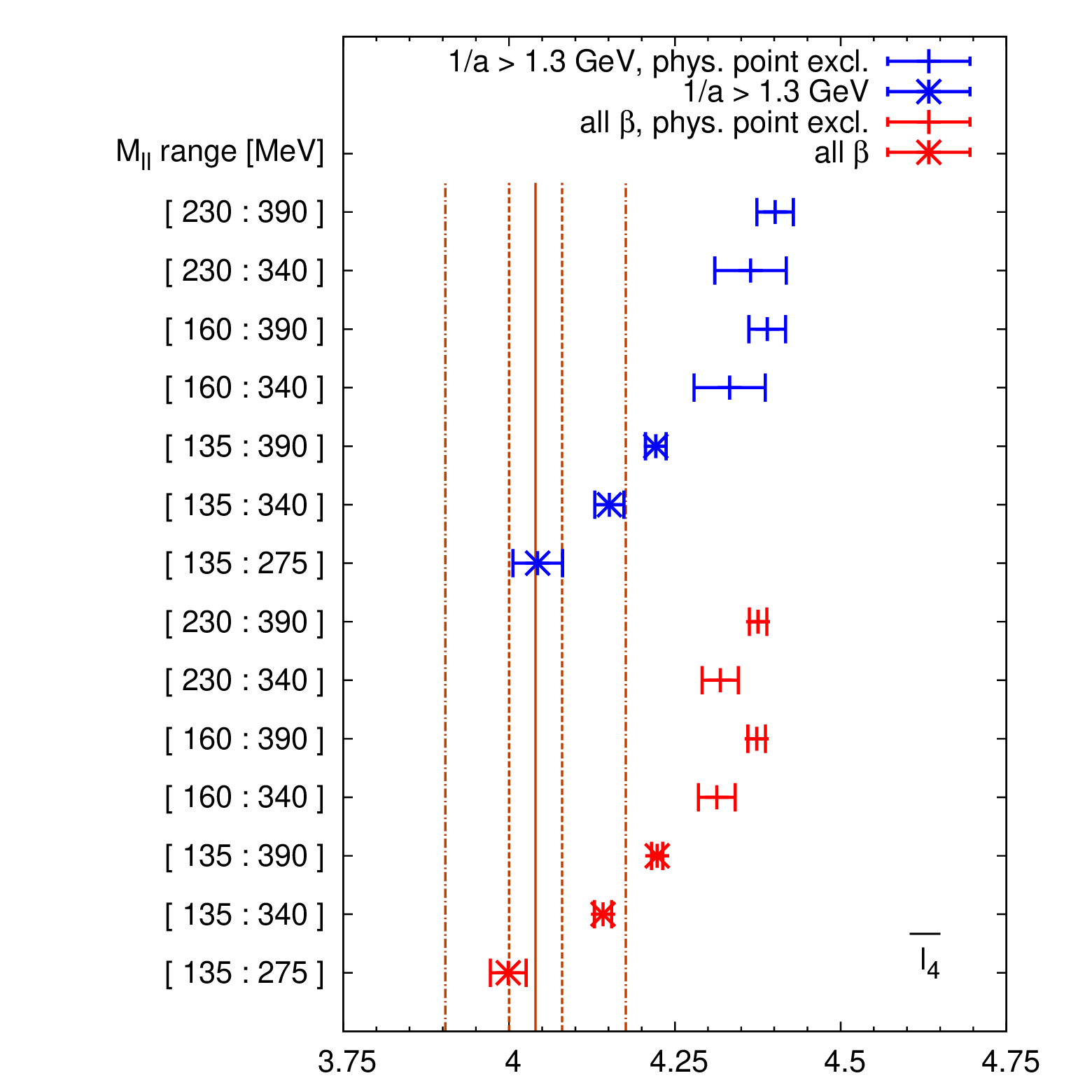}%
\includegraphics*[width=.33\textwidth]{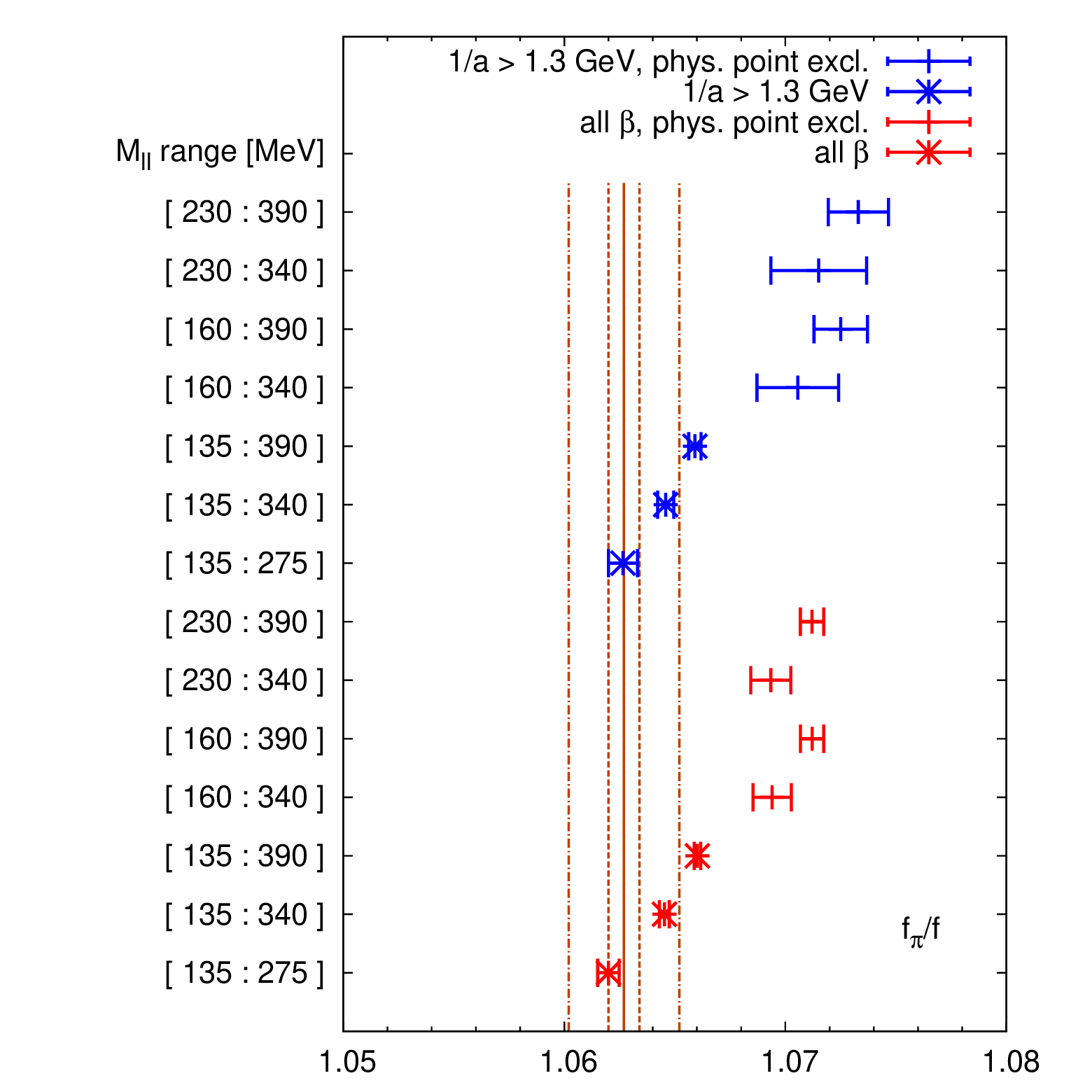}
\end{center}
\vspace*{\closercaption}
\caption{LECs obtained from NLO SU(2) chPT fits with different fit ranges: {\it left panel:} $\bar{l}_3$, {\it middle panel:} $\bar{l}_4$, {\it right panel:} $f_\pi/f$. {\it Blue points} denote fits where $1/a\geq1.35\,{\rm GeV}$, {\it red points} fits where all $\beta$ are included. Fits including the nearly physical points are marked by an {\it asterisk}. The {\it solid}, {\it dashed} and {\it dashed-dotted lines} display the central value, statistical and combined (stat.\ and syst.) error, resp., of our quoted results.}
\label{fig:summary_lbar3_lbar4_f}
\end{figure}

Here we are mainly interested in the SU(2) low-energy constants $\bar{l}_3$ and $\bar{l}_4$ which are related to the low-energy scales $\Lambda_3$, $\Lambda_4$, respectively. Therefore, in Fig.~\ref{fig:summary_lbar3_lbar4_f} we show the fitted values for these parameters obtained with different fit ranges. We also display the ratio $f_\pi/f$ as obtained from the various fits. Whereas for $\bar{l}_3$ (left panel), if at all, one could identify a shift in the result depending on whether or not the two coarsest lattices ensemble are excluded, for $\bar{l}_4$ and $f_\pi/f$ one observes a clear dependency on the fitted mass range, while the influence of excluding coarser lattice ensembles seems to have only a marginal effect. Eventually, we quote as our result for the low-energy constants the central value and statistical error obtained from the fit range $135\,{\rm MeV}\leq M_{ll}\leq275\,{\rm MeV}$,  $1/a\geq 1.3\,{\rm GeV}$ and take the variation with respect to that value from other fits including the nearly physical points (data marked by asterisks in Fig.~\ref{fig:summary_lbar3_lbar4_f}) as our estimate for the systematic error, so that we obtain:
\begin{equation}
\label{eq:LECs_fit}
 \bar{l}_3\;=\;2.90(11)_{\rm stat}(17)_{\rm syst}\,,\;\;\;\bar{l}_4\;=\;4.04(04)_{\rm stat}(13)_{\rm syst}\,,\;\;\;f_\pi/f\;=\;1.0627(07)_{\rm stat}(24)_{\rm syst}\,.
\end{equation}
%

%%%%%%%%%%%%%%%%%%%%%%%%%%%%%%%%%%%%%%%%%%%%%%%%%%%%%%%
% lbar3,4 and f from different fits
%%%%%%%%%%%%%%%%%%%%%%%%%%%%%%%%%%%%%%%%%%%%%%%%%%%%%%%
\vspace*{\afterFigure}
%%%%%%%%%%%%%%%%%%%%%%%%%%%%%%%%%%%%%%%%%%%%%%%%%%%%%%%

%%%%%%%%%%%%%%%%%%%%%%%%%%%%%%%%%%%%%%%%%%%%%%%%%%%%%%%
% NLO fits excl. the physical point
%
% only fPi, 1/a >= 1.35 GeV
%
% * 230 - 340 MeV
%
% * 230 - 390 MeV
%
%%%%%%%%%%%%%%%%%%%%%%%%%%%%%%%%%%%%%%%%%%%%%%%%%%%%%%%
\begin{figure}[t!]
\begin{center}
%%%
\includegraphics*[angle=-90,width=.49\textwidth]{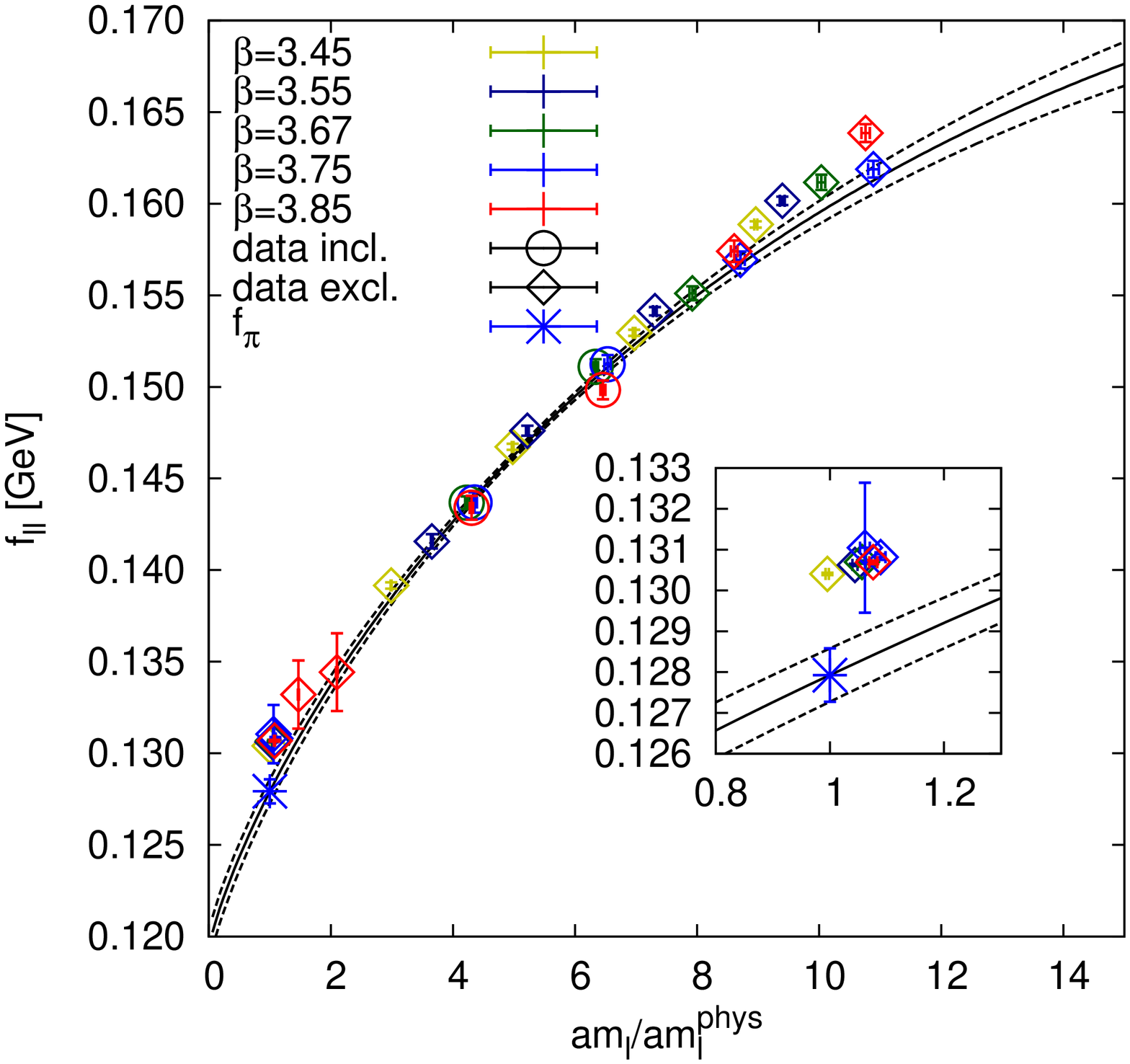}%
\includegraphics*[angle=-90,width=.49\textwidth]{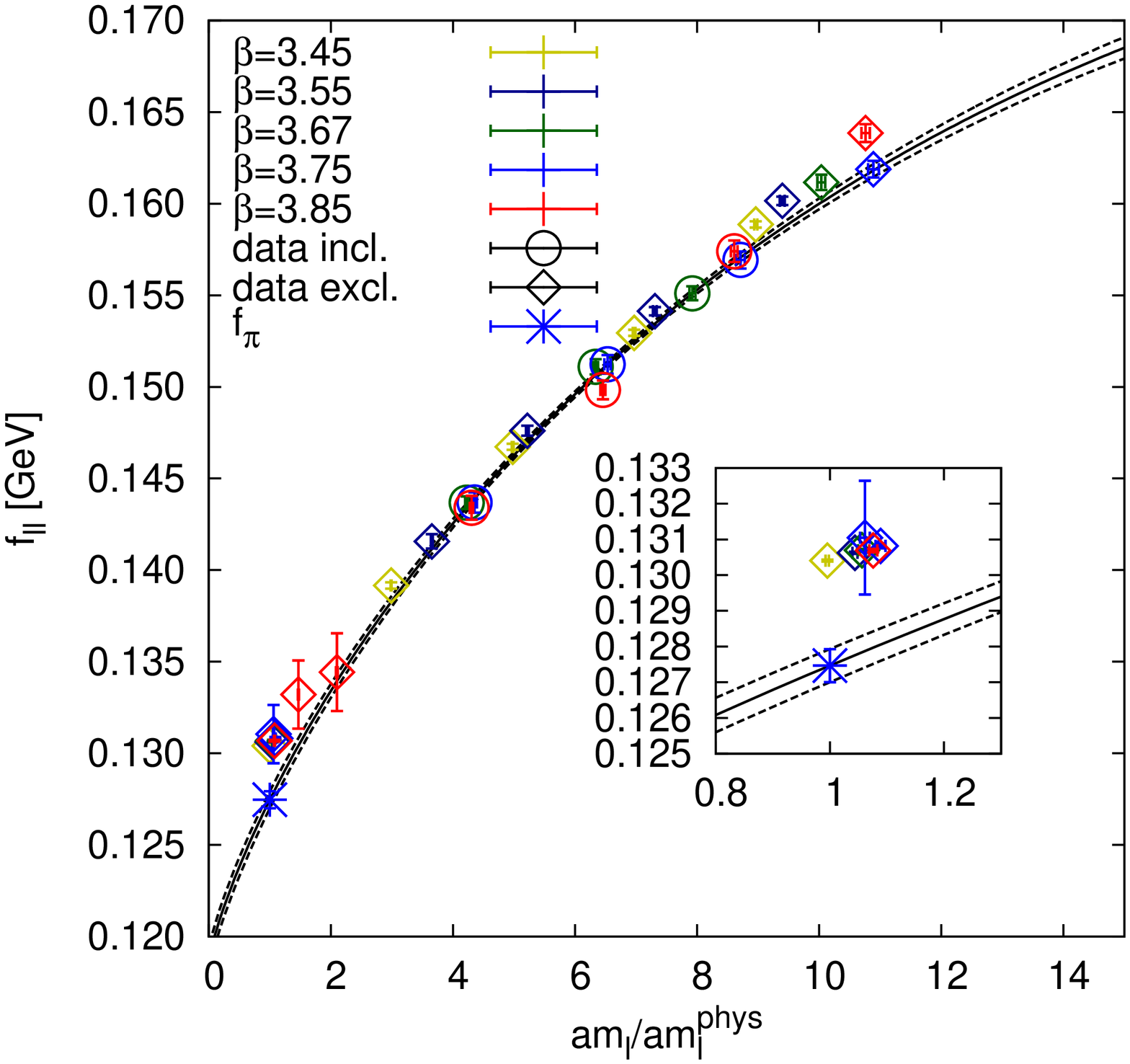}
%%%%
\end{center}
\vspace*{\closercaption}
\caption{Two examples for NLO SU(2) chPT fits excluding the nearly physical points (only $f_{ll}$ is shown here). {\it Left panel:} $230\,{\rm  MeV} \leq M_{ll} \leq 340\,{\rm MeV}$, {\it right panel:} $230\,{\rm  MeV} \leq M_{ll} \leq 390\,{\rm MeV}$; both $1/a\geq1.35\,{\rm GeV}$.}
\label{fig:fits_NLO_exclPhys}
\end{figure}
\vspace*{\afterFigure}
%%%%%%%%%%%%%%%%%%%%%%%%%%%%%%%%%%%%%%%%%%%%%%%%%%%%%%%

Since often lattice data from meson masses larger than the physical $M_\pi$ are extrapolated to the physical point using SU(2) chPT, we also investigated fit ranges excluding the physical point. In Fig.~\ref{fig:fits_NLO_exclPhys} the fits for the meson decay constant are shown for $230\,{\rm MeV}\leq M_{ll} \leq 340\,{\rm MeV}$ (left panel) and  $230\,{\rm MeV}\leq M_{ll} \leq 390\,{\rm MeV}$ (right panel). As one can see in the close-up view of the region near the physical point, the value for $f_\pi$ extrapolated from such a fit is below ($f_\pi^{\rm extr.}\approx 128(1)\,{\rm MeV}$) the values simulated near the physical point. As one can see from Fig.~\ref{fig:summary_lbar3_lbar4_f}, also $\bar{l}_4$ and $f_\pi/f$ are significantly changing, once the nearly physical points are excluded from the fit range.

%%%%%%%%%%%%%%%%%%%%%%%%%%%%%%%%%%%%%%%%%%%%%%%%%%%%%%%%%%%%%%%%%%%%%%%%%%%%%%%
\section{Fits to NNLO SU(2) chPT}
\label{subsec:fit.nnlo}
\vspace*{\closersection}

\begin{figure}[t!]
\begin{center}
%%%
\includegraphics*[angle=-90,width=.49\textwidth]{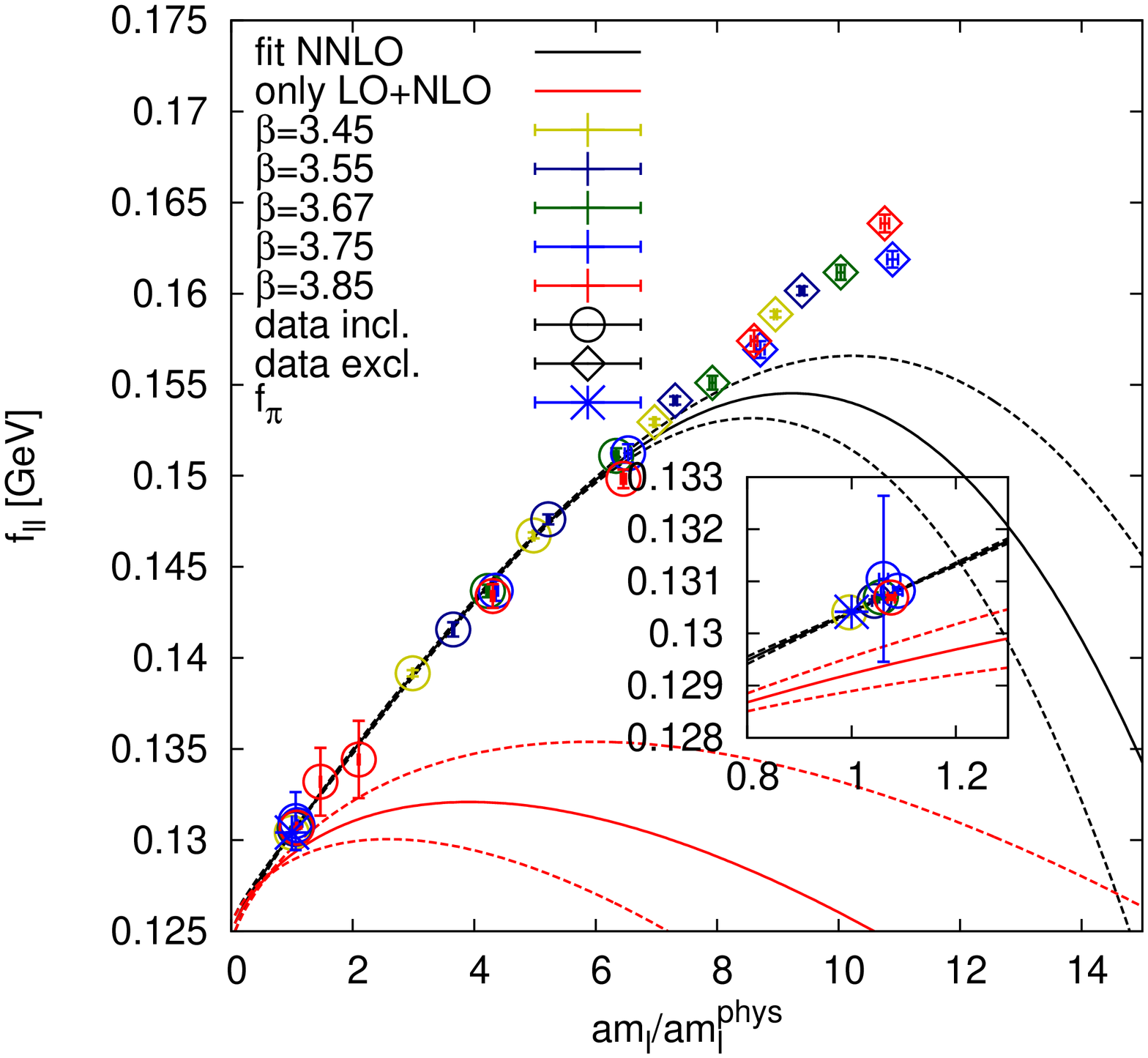}%
%
%\hspace{.05\textwidth}%
%
\includegraphics*[angle=-90,width=.49\textwidth]{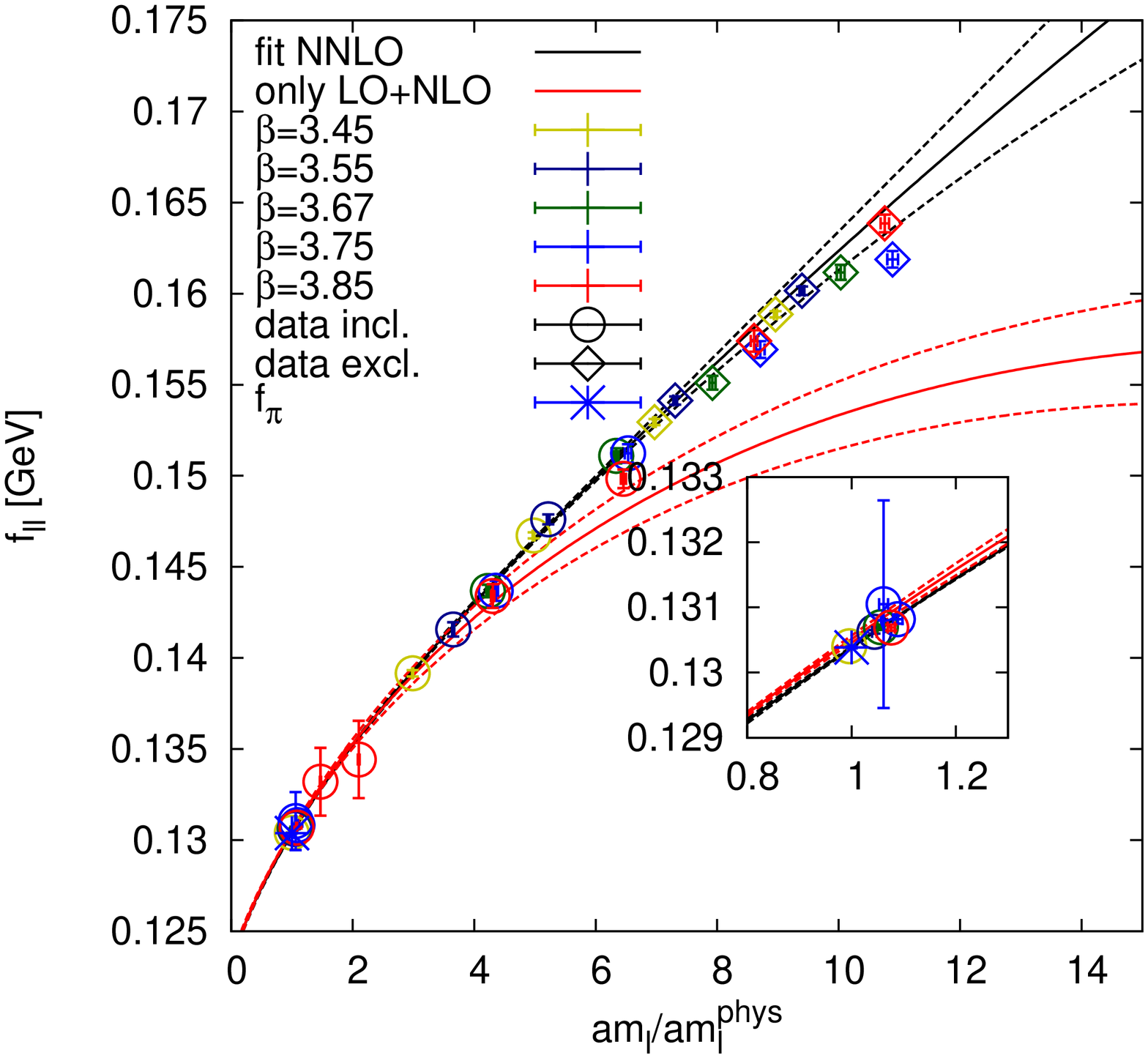}
%%%%
\end{center}
\vspace*{\closercaption}
\caption{Two examples for NNLO SU(2) chPT fits (only $f_{ll}$ is shown here). {\it Left panel:} without priors, {\it right panel:} using prior $\bar{l}_{12}=2.1\pm0.3$. Both use all $\beta$ and $135\,{\rm  MeV} \leq M_{ll} \leq 340\,{\rm MeV}$.}
\label{fig:fits_NNLO}
\end{figure}

Extending the SU(2) chPT fit formulae for the meson masses and decay constants to NNLO (e.g.\ cf.~\cite{Colangelo:2010et}), in our set-up three new fit parameters have to be added: a combination of the NLO low-energy constants $\bar{l}_1$, $\bar{l}_2$: $\bar{l}_{12}=(7\bar{l}_1+8\bar{l}_2)/15$ and two parameters for NNLO-LECs $k_m$, $k_f$. Again fitting our data for the meson masses and decay constants at various $\beta$ simultaneously now using the NNLO fit formulae without any constraints on the fit parameters (7 in total) leads to an unnatural order of the NLO- compared to the NNLO-contribution as can be seen from the left panel of Fig.~\ref{fig:fits_NNLO}. There the black line denotes the full fit up to NNLO and the red line only the contribution up to NLO, the large difference between the two being the NNLO-contribution. The situation can be improved  by using priors for some of the fit parameters, e.g., using a phenomenological estimate for $\bar{l}_{12}=2.1\pm0.3$ as can be obtained from values quoted for $\bar{l}_1$, $\bar{l}_2$ in \cite{Colangelo:2001df}. A fit using such a prior is shown in the right panel of Fig.~\ref{fig:fits_NNLO}, which describes the data well and has a reasonable ordering of the NLO- compared to NNLO-contribution. Still we refrain from using NNLO-chPT as long as we do not have enough data in the light quark mass region to constrain such fits without having to rely on additional input used for priors on the fit parameters. But it is reassuring to us, that by using such priors, a NNLO-fit results in NLO-LECs comparable to those found in our NLO-fits. 

%%%%%%%%%%%%%%%%%%%%%%%%%%%%%%%%%%%%%%%%%%%%%%%%%%%%%%%
%
%%%%%%%%%%%%%%%%%%%%%%%%%%%%%%%%%%%%%%%%%%%%%%%%%%%%%%%
\vspace*{\afterFigure}
%%%%%%%%%%%%%%%%%%%%%%%%%%%%%%%%%%%%%%%%%%%%%%%%%%%%%%%

%%%%%%%%%%%%%%%%%%%%%%%%%%%%%%%%%%%%%%%%%%%%%%%%%%%%%%%%%%%%%%%%%%%%%%%%%%%%%%%
\section{Conclusions}
\vspace*{\closersection}

From our NLO SU(2) chPT fits to meson masses and decay constants measured on staggered 2+1 flavor lattice simulations of QCD, we quote the following set of LECs (see Eq.~(\ref{eq:LECs_fit})) as our preliminary result: 

\[ \bar{l}_3\;=\;2.90\pm0.20\,,\;\;\;\bar{l}_4\;=\;4.04\pm0.14\,,\;\;\;f_\pi/f\;=\;1.0627\pm0.0025\,.\]

These values are in good agreement with other recent lattice determinations of LECs, for example the FLAG-report \cite{Colangelo:2010et} quotes $\bar{l}_3=3.2\pm0.8$ and $f_\pi/f=1.073(15)$ as lattice averages, while due to some tension in the results no value for $\bar{l}_4$ is quoted at the moment. Our findings also agree well with the phenomenological estimates $\bar{l}_3=2.9\pm2.4$ and $\bar{l}_4=4.4\pm0.2$ \cite{Colangelo:2001df} and $f_\pi/f=1.0719\pm0.0052$ \cite{Colangelo:2003hf,Colangelo:2010et}.

For a forthcoming publication we hope to have additional data points available at light quark masses corresponding to meson masses between 135 and 275 MeV. More details about our chiral fits will be reported there as well.

%\noindent{\small 
The speaker acknowledges support from the DFG SFB/TR 55 and the EU grant PITN-GA-2009-238353 (ITN STRONGnet).
%}

%%%%%%%%%%%%%%%%%%%%%%%%%%%%%%%%%%%%%%%%%%%%%%%%%%%%%%%%%%%%%%%%%%%%%%%%%%%%%%%

%%%%%%%%%%%%%%%%%%%%%%%%%%%%%%%%%%%%%%%%%%%%%%%%%%%%%%%%%%%%%%%%%%%%%%%%%%%%%%%
\vspace*{-.4cm}
\bibliography{references}
\bibliographystyle{h-physrev5}
%%%%%%%%%%%%%%%%%%%%%%%%%
%\begin{thebibliography}{99}
% \bibitem{...} ....
%\end{thebibliography}
%%%%%%%%%%%%%%%%%%%%%%%%%

%%%%%%%%%%%%%%%%%%%%%%%%%%%%%%%%%%%%%%%%%%%%%%%%%%%%%%%%%%%%%%%%%%%%%%%%%%%%%%%
\end{document}